\documentclass[twocolumn,prb,superscriptaddress]{revtex4-2}
\usepackage[utf8]{inputenc}
\usepackage{color}
\usepackage{amsmath}
\usepackage{graphicx}
\usepackage{siunitx}
\makeatletter

\providecommand{\tabularnewline}{\\}

\usepackage{multirow}
\usepackage[normalem]{ulem}

\usepackage{hyperref}

\makeatother

\begin{document}
\title{Influence of native defects on magneto-optoelectronic properties of $\alpha$-MoO$_{3}$ }
\author{Poonam Sharma}
\email{pspoonamsharma44@gmail.com}
\affiliation{Department of Physics, Indian Institute of Technology Bombay, Mumbai
400076, India}
\author{Vikash Mishra}
\email{vikash2035@gmail.com}
\affiliation{Department of Physics, Indian Institute of Technology Bombay, Mumbai
400076, India}
\author{Alok Shukla}
\email{shukla@iitb.ac.in}

\affiliation{Department of Physics, Indian Institute of Technology Bombay, Mumbai
400076, India}
\begin{abstract}
Semiconducting oxides possess a variety of intriguing electronic,
optical, and magnetic properties, and native defects play a crucial
role in these systems. In this study, we study the influence of native
defects on these properties of $\alpha$-MoO$_{3}$ using the first-principles
density functional theory (DFT) calculations. From the formation energy
calculations, it is concluded that Mo vacancies are difficult to form
in the system, while O and Mo-O co-vacancies are energetically quite
favorable. We further find that vacancies give rise to mid-gap states
(trap states) that remarkably affect the magneto-optoelectronic properties
of the material. Our calculations indicate that a single Mo vacancy
leads to half-metallic behavior, and also induces a large magnetic
moment of 5.98~$\mu_{B}$. On the other hand, for the single O vacancy
case, the band gap disappears completely, but the system remains in
a non-magnetic state. For Mo-O co-vacancies of two types considered
in this work, a reduced band gap is found, along with an induced magnetic
moment of 2.0~$\mu_{B}$. Furthermore, a few finite peaks below the
main band edge are observed in the absorption spectra of configurations
with Mo and O vacancies, while they are absent in the Mo-O co-vacancies
of both types, just like in the pristine state. From the ab-initio
molecular dynamics simulations, stability and sustainability of induced
magnetic moment at room temperate is verified. Our findings will enable
the development of defect strategies that maximize the functionality
of the system, and further help in designing highly efficient magneto-optoelectronic
and spintronic devices. 
\end{abstract}
\maketitle
\section{Introduction}

In recent years, transition metal oxides have been intensely investigated
by the research community due to their extensive usage in various
fields such as field emission devices~\cite{wei2009synthesis},
storage devices~\cite{chernova2009layered,saji2012molybdenum,lee2008reversible},
light-emitting diodes~\cite{you2007improved,meyer2012transition,bao2010electronic}, high-power transitiors, and other optoelectronic devices. In particular, molybdenum trioxide (MoO$_{3}$), which is an n-type semiconductor, is a well-known transition metal oxide owing to its high dielectric constant, high work function (i.e.,~$\sim$~6.8 eV), and tunable semiconducting characteristics~\cite{de2017molybdenum,meyer2012transition,lin2010electrochromic}. The layers of this material are easily peeled away from the bulk crystal because of the weak van der Waals (vdW) interaction between them. Moreover, due to having effective hole transport in $\alpha$-MoO$_{3}$, it is a good candidate for use in photodetectors~\cite{xiang2014gap}, field-effect transistors~\cite{balendhran2013enhanced}, field-effect biosensors~\cite{balendhran2013field}, gas sensors~\cite{ji20162d}, resistive memory devices~\cite{rahman2018reversible}, electrochromic and photochromic devices~\cite{saji2012molybdenum}, and supercapacitor applications~\cite{hanlon2014production}. It has been observed that MoO$_{3}$ typically crystallizes in
three phases, namely $\alpha$-MoO$_{3}$ (orthorhombic structure),
$\beta$-MoO$_{3}$ (monoclinic structure), and hexagonal MoO$_{3}$.
Octahedron composed of MoO$_{6}$ provides a primary framework for
synthesizing different phases of MoO$_{3}$. Double layers of octahedral
MoO$_{6}$ connect with van der Waals (vdW) interaction to form the layered orthorhombic
phase of $\alpha$-MoO$_{3}$. In the past, $\alpha$-MoO$_{3}$ has
been synthesized using various experimental techniques \citep{gao2012hydrothermal,boudaoud2006structural,chithambararaj2011hydrothermal,sinaim2012free,sreedhara2013synthesis,patzke2004one},
and studies reveal that $\alpha$-MoO$_{3}$ with the orthorhombic
structure is the most stable phase.

Measurements show that the experimental band gap of $\alpha$-MoO$_{3}$
lies in the range of 3.0-3.3 eV~\citep{chen2010single,hu2009moo3,carcia1987synthesis}.
Huang~\emph{et al.} performed the density functional theory (DFT)
calculations, including vdW interactions, and obtained an
indirect band gap of 1.62 eV~\citep{huang2014impact}. Li~\emph{et
al.} performed PBE+D2 calculations for the single-layer and bulk phase
of MoO$_{3}$ and predicted an indirect band gap of 1.73 eV and 1.71
eV, respectively~\citep{li2013tuning}. Akande~\emph{et al.} performed
the GGA+U calculation by considering two different U values. By applying
U = 4.3 and 6.3 eV on Mo($d$) orbitals, they observed a good agreement
of lattice constants with experimental values, but the band gap was
found to be close to 2 eV for both values of U~\citep{akande2016vacancy}.
Inzani~\emph{et al.} investigated the influence of van der Waals
dispersion effects, and also included several U corrections on Mo($d$)
orbitals~\citep{inzani2016van}. They concluded that, while the dispersion
correction improves the description of the crystal structure significantly,
a change in U value applied on Mo ($d$) orbitals from 2 to 8 eV has
essentially no influence on the band gap. Das~\emph{et al.} also
studied the properties of MoO$_{3}$ at the DFT+U/D$_{3}$ and DFT+U/D$_{3}$/SO levels~\citep{das2019structural}
of theory by applying U = 5 eV on Mo($d$) orbitals, and observed
a decrement in the band gap to 1.65 eV, as compared to PBE value of
1.99 eV. Interestingly, the authors also applied U corrections on
the delocalized O 2$p$ orbitals and obtained much-improved results
on the properties such as the indirect band gap of the system, lattice
parameters, formation enthalpy, static dielectric constant, and dissociation
energy of O$_{2}$ were computed accurately. The authors concluded
that applying U correction in the range 3 to 7 eV on Mo($d$) orbitals
results in worsening of the band gap values and recommended a U
correction only on O 2$p$ orbitals while studying other properties
of bulk MoO$_{3}$. In all the studies discussed above, the computed
band gap is well below the experimental value. However, Peelaers~\emph{et
al.}~\citep{peelaers2017controlling} performed hybrid functional
(HSE06) based calculations to match the band gap value with the experimental
data. Their research shows an indirect band gap of 3.19 eV. Qu~\emph{et
al.} reported a band gap of 2.884 eV~\citep{qu2017electronic}, and
Akande~\emph{et al.} reported a band gap of 3.1 eV~\citep{akande2016vacancy}
using the HSE06 functional.

Additionally, a lot of work has been done to narrow the band gap of $\alpha$-MoO$_{3}$ in order to exploit it for solar energy applications~\cite{bandaru2018tweaking}. In the pure polycrystalline material, it is well established
that there will be a significant number of defects and impurities
in the system~\citep{mishra2019investigation,mishra2018diffuse}.
Even the presence of a small number of defects in the crystal geometry
influences the magneto-optoelectronic properties of the material in
a non-trivial manner, as a result, many researchers have focused on
understanding the performance of devices in the presence of lattice
imperfections. The behavior of defects largely depends on the nature
of the material. For example, in oxide materials such as MgO, the
excess number of electrons introduced into the system due to oxygen
defects are localized on its surface, whereas in the case of TiO$_{2}$,
electrons are delocalized~\citep{jupille2015defects}. Photoluminescence
spectroscopy~\citep{kan2005blue}, Raman spectroscopy~\citep{wu2010probing},
valence band spectroscopy~\citep{jupille2015defects}, and first-principle
calculations studies~\citep{courths1989bulk,goes2018identification,deak2019optimized}
have been widely used to probe the signature of defects in oxide
systems. Defects introduce some additional states (trap states) in
the band gap region (between the valence and conduction band) that
drastically change the magneto-optoelectronic properties of the system.
For better device performance and other applications, there is obviously
a need to understand not just the origin of defects, but also their
influence on other properties. Many researchers have employed the
first-principles DFT-based calculations for this purpose and computed
not just the magneto-optoelectronic properties in the presence of
defects, but also the stability of the native defects in semiconducting
oxides~\citep{freysoldt2014first}.


Theoretically and experimentally, previously vacancy-induced magnetism studies in oxide materials, such as ZnO, Y$_2$O$_3$, TiO$_2$, SnO$_2$, HfO$_2$, etc., have been extensively performed   ~\cite{kohan2000first,zheng2006native,venkatesan2004unexpected,pemmaraju2005ferromagnetism,elfimov2002possible,rahman2008vacancy,rahman2013stabilizing,khalid2009defect}. For example, Elfimov~\emph{et al.} performed DFT calculations on CaO and found that a single Ca vacancy induced a magnetic moment of 2~$\mu_{B}$ at 3.125 at$\%$ of vacancy concentration, 88$\%$ contribution to the magnetic moment comes from the O ions, i.e., nearest neighbors of Ca vacancy, and the system becomes half-metallic ferromagnetic~\cite{elfimov2002possible}. In the O vacancy, the system still shows non-magnetic behavior. Pemmaraju~\emph{et al.} performed the DFT study on HfO$_2$ by considering a $2\times2\times2$ supercell having 96 atoms and reported that a single Hf vacancy (i.e., 3.125 at$\%$) induced a magnetic moment of 3.5~$\mu_{B}$, while O vacancy (i.e., 1.562 at$\%$) does not induce any magnetic moment in the system~\cite{pemmaraju2005ferromagnetism}. Khalid~\emph{et al.} reported from Positron annihilation spectroscopy measurements and further confirmed by DFT calculations that Zn vacancies are responsible for inducing magnetism in ZnO~\cite{khalid2009defect}. Rahman~\emph{et al.} studied the defects-induced magnetism in SnO$_2$ and Li-doped SnO$_2$~\cite{rahman2013stabilizing,rahman2008vacancy}. In defects induced magnetism study of SnO$_2$, the authors reported that a single Sn vacancy at the vacancy concentration of 6.25 at$\%$ induced a magnetic moment of 4.0~$\mu_{B}$, where O atoms surrounded by the Sn vacancy defect contribute the most to the magnetic moment. In the case of a single O vacancy, the system still remains non-magnetic. Further, Pandey~\emph{et al.} studied the impact of Sr, Ti, and O vacancy defects on the electronic and optical properties of TiO$_2$ and SrTiO$_3$~\cite{pandey2022elucidating}. From the experimental measurements, the authors found that vacancy defects enhance optoelectronic properties, such as photocurrent, optical absorption, etc. Wang~\emph{et al.} studied the effect of O vacancy defects on the electronic and optical properties through Raman spectroscopy, photoluminescence spectroscopy, absorption spectroscopy, and photoemission spectroscopy in the V$_2$O$_5$ nanowires~\citep{wang2016defect}. Due to O vacancy defects, an increase in the optical band gap from 1.95 eV to 2.45 eV was found. We believe that similar to these works on other transition metal oxides, a study of the magnetic and optical properties of defective $\alpha$-MoO$_{3}$ is also worthwhile.

Therefore, in this work, we undertake a systematic study of the variation of magneto-optoelectronic properties of $\alpha$-MoO$_{3}$ with different defects (vacancies) and also to understand the stability of the system in response to these defects. In most previous studies of vacancies in $\alpha$-MoO$_{3}$, predominantly O vacancies were considered~\cite{inzani2016van,peelaers2017controlling,guo2014origin,lambert2017formation}. In the present work, along with the O vacancy, the Mo vacancy, and Mo-O co-vacancies have also been considered. A systematic study using first-principles DFT calculations on pristine and self-deficient~$\alpha$-MoO$_{3}$ has been carried out, and the obtained results are compared with existing experimental and theoretical results.

\section{COMPUTATIONAL DETAILS}

Spin-polarized DFT calculations were carried
out using the Vienna Ab-initio Simulation Package (VASP) code~\citep{kresse1996efficient,kresse1996efficiency}
in conjunction with the projector augmented wave (PAW) pseudopotentials
method~\citep{mortensen2005real,kresse1999ultrasoft}. The generalized
gradient approximation (GGA), as implemented in the Perdew-Burke-Ernzerhof
(PBE) functional, was employed in the calculations~\citep{perdew1996generalized}.
In order to benchmark our GGA results, we also repeated some of our
calculations using HSE06 functional~\citep{qu2017electronic}. Convergence
tests were carried out systematically for the plane-wave cutoff energy and for the number of k-point used in the sampling of the Brillouin
Zone (BZ). A plane-wave basis set with an energy cutoff of 500 eV
has been used throughout the calculations. For BZ sampling, Monkhorst-pack~\citep{monkhorst1976special}
mesh with a k-point grid $5\times10\times9$ was used for geometry
optimization of the unit cell, and $7\times12\times11$ for the density
of states calculations. The unit cell of $\alpha$-MoO$_{3}$ contains
16 atoms. The valence electronic configurations of Mo and O atoms
for the PAW pseudopotentials were taken as 4$p^{6}$5$s^{1}$4$d^{5}$,
and 2$s^{2}$2$p^{4}$, respectively. We optimized the atomic coordinates
along with the lattice parameters of the pristine $\alpha$-MoO$_{3}$
unit cell. The convergence criterion of the Hellman-Feynman force
on all atoms was set to 5$\times$10$^{-3}$ eV, while the energy
threshold was fixed at 10$^{-5}$ eV. \emph{Ab-initio} molecular dynamics
(AIMD) simulations were performed at 300 K to check the stability
and sustainability of the defects-induced magnetic moment using canonical
ensemble (NVT) for a time period of 5 ps, with the step size of 1
fs.

To study the optical properties of the system, we have computed the
complex dielectric function and, subsequently, the absorption coefficients,
using Ehrenreich and Cohen's~\citep{ehrenreich1959self} formalism.
The frequency ($\omega$) dependent complex dielectric function $\epsilon$($\omega$)
is given as 
\begin{equation}
\epsilon(\omega)=\epsilon{}_{1}(\omega)+i\epsilon{}_{2}(\omega),
\end{equation}
where, $\epsilon{}_{1}$($\omega$) and $\epsilon{}_{2}$($\omega$)
denote the real and imaginary part of the dielectric function, with
$\epsilon{}_{2}$($\omega$) defined as
\begin{equation}
\epsilon{}_{2}(\omega)=\frac{e^{2}\hbar}{\pi m^{2}\omega^{2}}\sum_{v,c}\int_{BZ}\lvert\langle u_{c{\bf k}}\lvert\hat{{\bf e}}\cdot\nabla\rvert u_{v{\bf k}}\rangle\rvert^{2}\delta[\omega_{cv}({\bf k})-\omega]d^{3}k.
\end{equation}
Above, the matrix element $\langle u_{c{\bf k}}\lvert\hat{{\bf e}}\cdot\nabla\rvert u_{v{\bf k}}\rangle$
corresponds to the direct transitions between the valence and the
conduction states, $\hat{{\bf e}}$ represents the polarization vector
of the incident photon, $\omega_{cv}({\bf k})$= $E_{c{\bf k}}$-$E_{v{\bf k}}$
denotes the excitation energy, and $u_{c{\bf k}}$ and $u_{v{\bf k}}$
are the periodic Bloch wave functions of wave vector ${\bf k}$ for
the conduction and valence states, respectively. Moreover, the Kramers-Kronig
relations can be used for the calculation of the real part $\epsilon_{1}(\omega)$
from $\epsilon{}_{2}$($\omega$)
\begin{equation}
\epsilon_{1}(\omega)=\frac{2}{\pi}p\int_{0}^{\infty}\frac{\omega'\epsilon{}_{2}(\omega')}{\omega^{'2}-\omega^{2}}d\omega',
\end{equation}
where $p$ denotes the principle value of the integral. The two dielectric
functions are then used to calculate the reflectivity, absorption
coefficient, refractive index, electron energy-loss spectrum, and
other constants~\citep{ravindran1999electronic}. In this work, we
compute and analyze the absorption coefficient $\alpha(\omega)$ given
as 
\begin{equation}
\alpha(\omega)=\frac{\sqrt{2}\,\omega}{c}\left(\sqrt{(\epsilon{}_{1}(\omega))^{2}+(\epsilon{}_{2}(\omega))^{2}}-\epsilon{}_{1}(\omega)\right)^{\frac{1}{2}}\textrm{.}
\end{equation}
Moreover, it is necessary to know the formation energies when studying
the effect of defects or impurities on the electronic structure because
they describe the stability of the defective systems. We have used
the following equation (Eq.~\ref{eq:eq5}) for the calculation of
the formation energies of systems with defects 
\begin{equation}
E_{form}=E_{defect}-E_{prisitne}+\sum_{k}\Delta n_{k}u_{k}+q(E_{VBM}+E_{F}).\label{eq:eq5}
\end{equation}

Above, $E_{defect}$ is the energy of the defective system, whereas
$E_{pristine}$ denotes the energy of the pristine system. Furthermore, ${\Delta}n_{k}$
is the number of $k$ types of atoms that are removed from the system
and $u{_{k}}$ is their chemical potential. The exchange of electrons in charged systems with charge $q$ is represented by the formula $q$($E_{VBM}$ + $E_{F}$), where $E_{F}$ is the Fermi level, and $E_{VBM}$ shows the valence-band maximum energy in the pristine system.

\section{RESULTS AND DISCUSSION }

\subsection{Structure Analysis}

\begin{figure}[ht]
\includegraphics[width=1\linewidth]{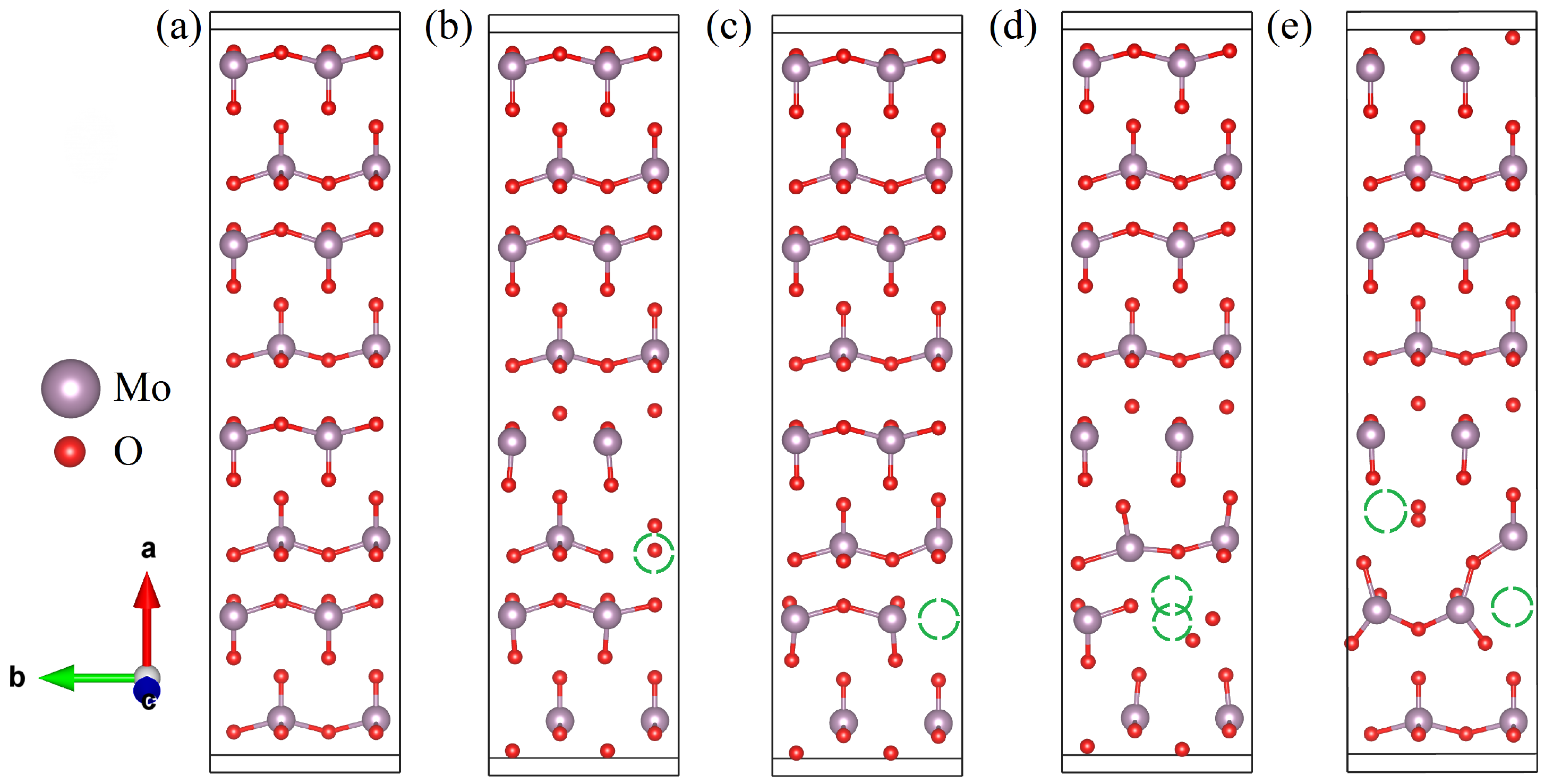}\caption{The optimized 2$\times$2$\times$1 supercell structures of $\alpha$-MoO$_{3}$:
(a) pristine structure, (b) with a single Mo vacancy, (c) with a single
O vacancy, and (d,e) with Mo-O co-vacancies of types P1 and P2, respectively.
Small green circles show the vacant positions of the respective atoms.}

\label{fig:Fig1.1} 
\end{figure}

In this work we study the orthorhombic phase of molybdenum trioxide
($\alpha$-MoO$_{3}$) with the space group symmetry Pnma (62). Lattice
parameters of the pristine $\alpha$-MoO$_{3}$ unit cell after performing
geometry optimization were found to be $a=\SI{14.80}{\angstrom}$, $b=\SI{3.72}{\angstrom}$, $c=\SI{3.97}{\angstrom}$~\citep{akande2016vacancy},
and in Fig.~\ref{fig:Fig1.1}(a) we show the optimized structure
for its $2\times2\times1$ supercell.\textcolor{red}{{} }

To study the variation of magneto-opto-electronic properties for various
types of defects (different types of vacancies) are created in $2\times2\times1$
supercell. Figs.~\ref{fig:Fig1.1}(b)-\ref{fig:Fig1.1}(e) show the
optimized $2\times2\times1$ supercell structure of $\alpha$-MoO$_{3}$
with Mo vacancy (V$_{Mo}$), O vacancy (V$_{o}$), and Mo-O co-vacancies
(V$_{Mo+O}$) of types P1 and P2, respectively.

\subsection{Band Structure and Density of States}

\subsubsection{Pristine Material}

\begin{figure}[ht]
\includegraphics[width=1\linewidth]{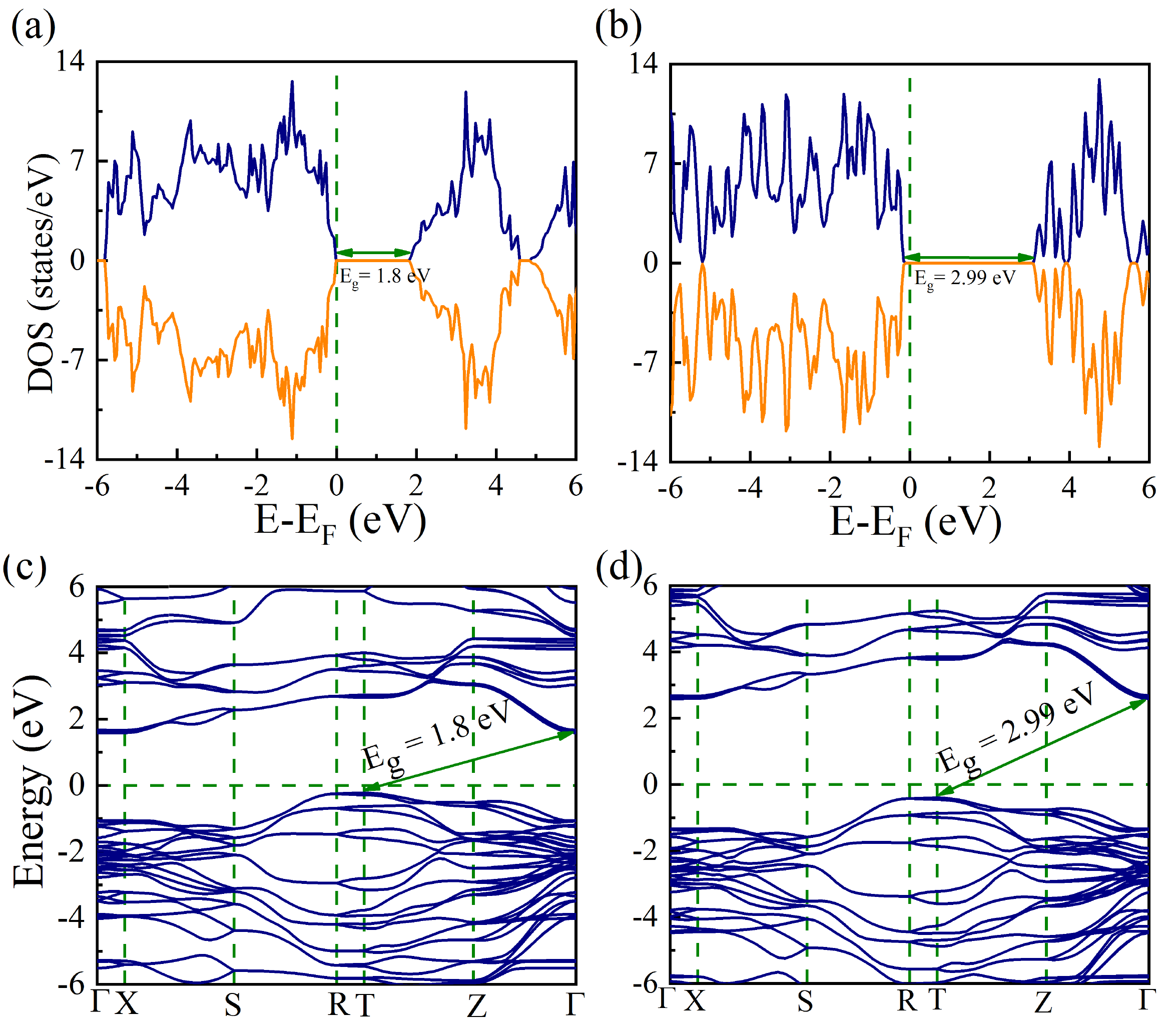}\caption{(a,b) The total density of states and (c,d) electronic band structure
of pristine unit cell of $\alpha$-MoO$_{3}$ using PBE and HSE06
functional, respectively. }
\label{fig:Fig2.1} 
\end{figure}

\begin{figure}[ht]
\includegraphics[width=1\linewidth]{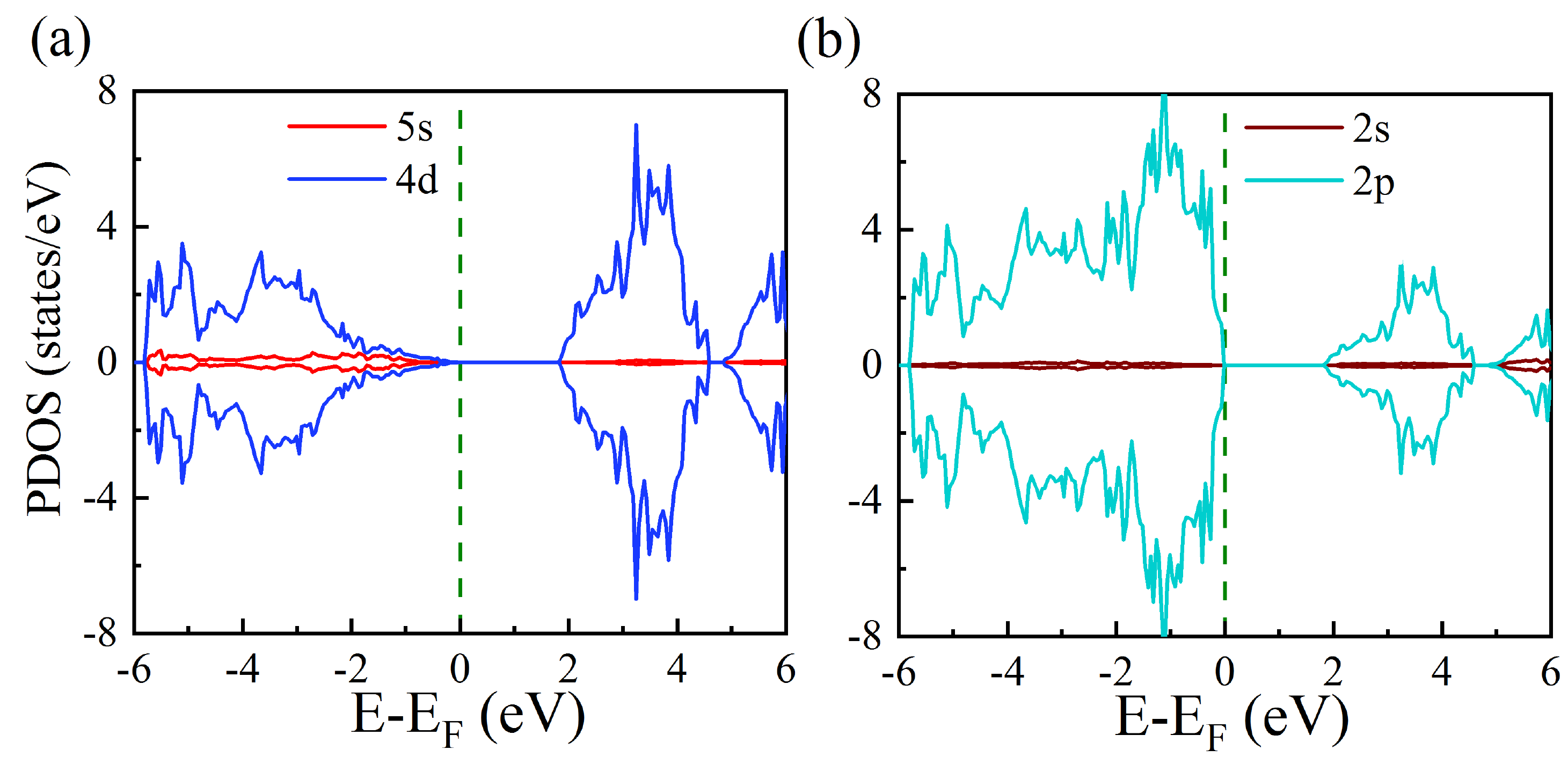}\caption{Partial density of states of (a) Mo (5$s$ and 4$d$ orbitals) and
(b) O (2$s$ and 2$p$ orbitals) atoms using GGA.}
\label{fig:Fig3.1} 
\end{figure}

Firstly, we studied the ground-state electronic properties of $\alpha$-MoO$_{3}$
using spin-polarized DFT simulations. The total density of states
(TDOS) and band structure of pristine $\alpha$-MoO$_{3}$ calculated
using the GGA and HSE06 functionals are presented in Figs. \ref{fig:Fig2.1}(a)-\ref{fig:Fig2.1}(d).
The GGA-level TDOS plot (see Figs.~\ref{fig:Fig2.1}(a)) clearly
shows symmetry between the up and down spin states indicating non-magnetic
behavior, and a finite band gap. As shown in the PBE band structure
plot (Fig. \ref{fig:Fig2.1}(c)), we obtain an indirect band gap of
1.8 eV~\citep{qu2017electronic}. Along the high-symmetry directions
corresponding to $\Gamma$-X and R-T, almost degenerate eigenvalues
are obtained in the valence and conduction bands.
Furthermore, the valence band maxima (VBM) and conduction band minima
(CBM) are positioned at T and $\Gamma$ points, respectively. Our
results on the band structure are in good agreement with the previously
reported results~\citep{qu2017electronic,dandogbessi2016first}.
In addition, we performed GGA+U calculations to match our band gap
with the experimental value. 
We first applied U correction in the range 2 to 12 eV, only on Mo(4$d$)) electrons. However, the maximum value of the calculated band gap was 2.54 eV for U=9.0 eV, which is still not in good agreement with the experimental value of 3.0-3.3 eV~\citep{chen2010single,hu2009moo3,carcia1987synthesis}. Given the fact that PDOS (see Fig.~\ref{fig:Fig3.1}) analysis reveals that the VBM is entirely composed of O 2$p$ orbitals, thus, next we applied the U correction also to the O(2$p$) electrons, in addition to Mo(4$d$) ones. We found that the choice U (2$p$) = 7 eV for O and $8 \leq U(4d) \leq 12$ eV for Mo does improve the band gap in the sense that it is closer to the experimental value. However, the calculations predict the band gap of the minority spin to be significantly larger than that of the majority spin. Similar behavior is also found in the case i.e., U $>$ 9 eV on Mo(4$d$) electrons. This result is in complete contradiction with our GGA and HSE06 results, therefore, we conclude that the DFT+U approach is yielding erroneous results for this system. Therefore, we repeated our calculations using a more accurate hybrid
functional, i.e., HSE06, and the results are presented in Figs. \ref{fig:Fig2.1}(b)
and \ref{fig:Fig2.1}(d). Indeed, we found an indirect band gap of
2.99 eV using the HSE06 functional, which is in excellent agreement
with the experimental value~\citep{sinaim2012free}. But, we note
that the band structure computed using the HSE06 functional is quite
similar to that obtained from GGA, except for the increased band gap.

Further, to achieve a better understanding of the contribution of different
orbitals in the TDOS, we have calculated the projected density of
states (PDOS) of Mo and O atoms, and presented it in Figs. \ref{fig:Fig3.1}(a)
and \ref{fig:Fig3.1}(b), respectively. We have computed the PDOS
for the 5$s$ and 4$d$ orbitals of the Mo atom, along with the 2$s$
and 2$p$ orbitals of the O atom. From the figure, it is obvious that
the majority of the contribution to the TDOS comes from the 4$d$
orbitals of Mo and 2$p$ orbitals of the O atom. Furthermore, the
dominant contribution to the TDOS near the Fermi energy is obtained
from the 2$p$ orbital of the O atom. 

\subsubsection{Defective Material}

Next, we studied the vacancy-induced electronic properties of $\alpha$-MoO$_{3}$,
and for this purpose, we first created Mo, O, and Mo-O co-vacancies
of types P1 and P2 in the $2\times2\times1$ supercell, followed by
geometry optimization. From the optimized geometry in each case, the
band structure, DOS, and optical absorption spectrum were computed.
We computed the optical absorption spectra because, experimentally,
defect signatures are successfully investigated in oxide samples using
optical absorption spectroscopy~\citep{du2010multifunctional,mishra2018diffuse,mishra2019investigation}.
Furthermore, DOS can be used for interpreting the experiments, which
are performed using transmission electron microscopy~\citep{hashimoto2004direct}
and scanning tunneling microscopy~\citep{ugeda2010missing}, thus revealing
information on such intrinsic defects in the system. In bulk systems,
defect states become very sensitive to the variation in the position
of defects in the supercell. The purpose of considering two types
of co-vacancies in the supercell is to see whether, or not, the magneto-optoelectronic
properties change in the oxides samples with the relative locations
of the vacancies.
\begin{figure}[ht]
\includegraphics[width=1\linewidth]{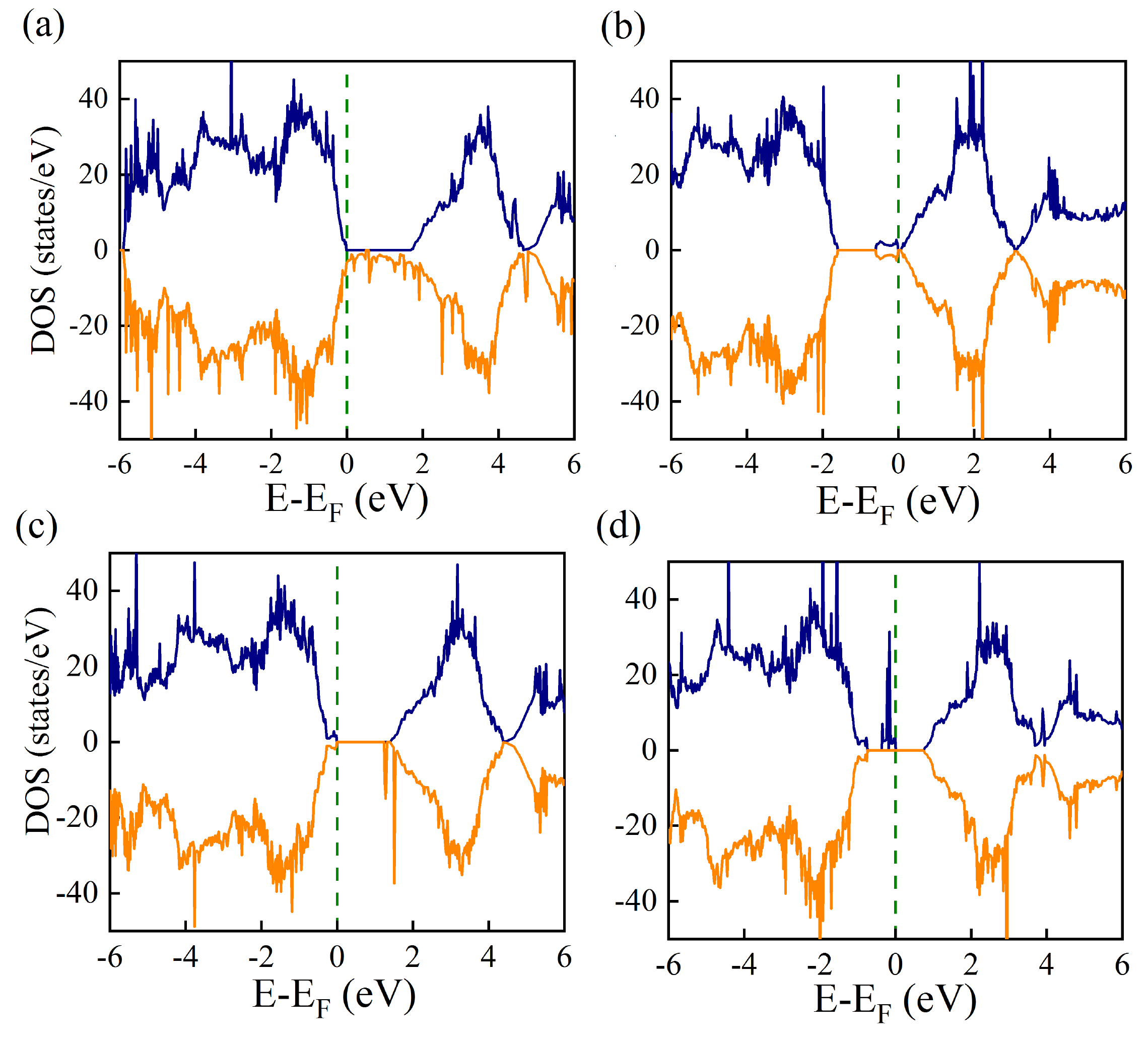}\caption{The total density states of $\alpha$-MoO$_{3}$ (2$\times$2$\times$1
supercell) with (a) Mo vacancy, (b) O vacancy, and Mo-O co-vacancies
of types (c) P1 and (d) P2, respectively, using GGA.}
\label{fig:Fig4.1} 
\end{figure}

\begin{figure*}
\includegraphics[width=1\linewidth]{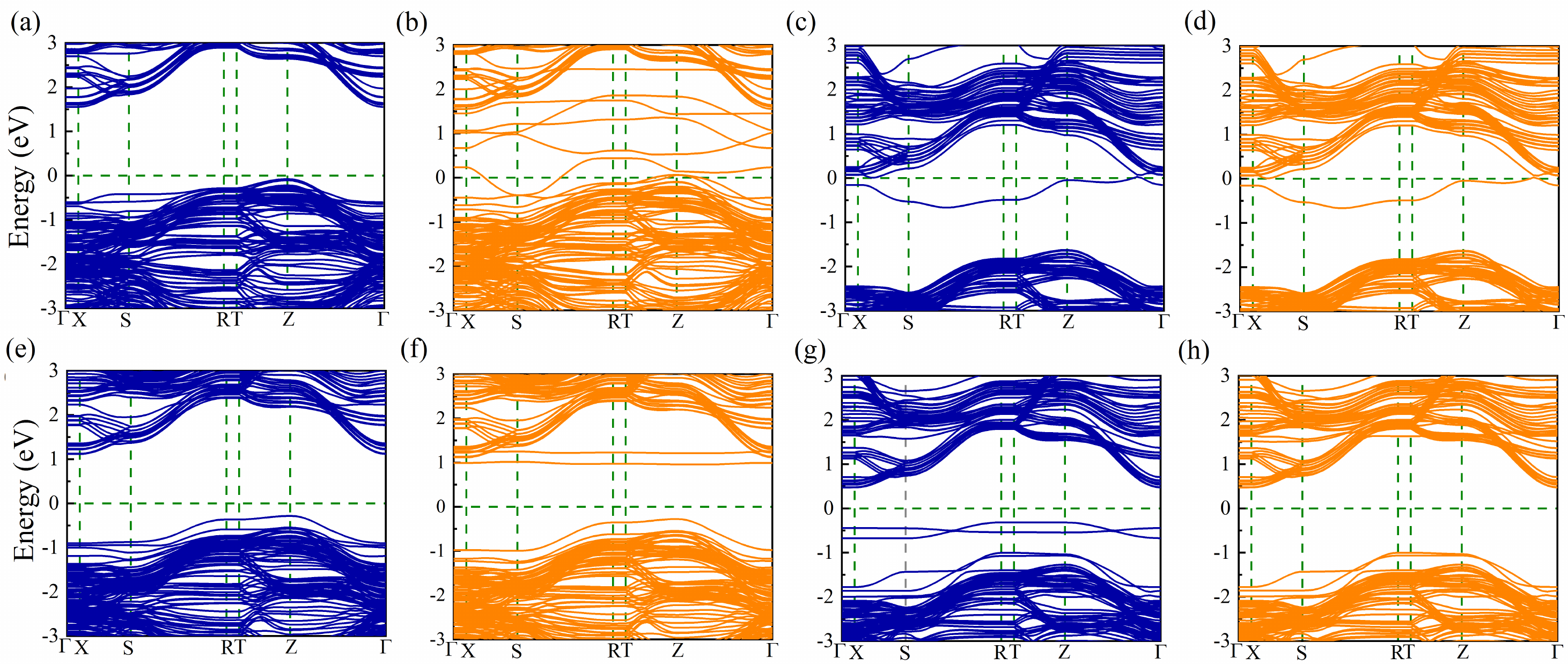}\caption{Electronic band structure of $\alpha$-MoO$_{3}$ (2$\times$2$\times$1
supercell) with (a,b) Mo vacancy, (c,d) O vacancy, and Mo-O co-vacancies
of types (e,f) P1 and (g,h) P2 for the spin-up and spin-down states,
respectively.}
\label{fig:Fig5.1} 
\end{figure*}

\begin{figure}[ht]
\includegraphics[width=1\linewidth]{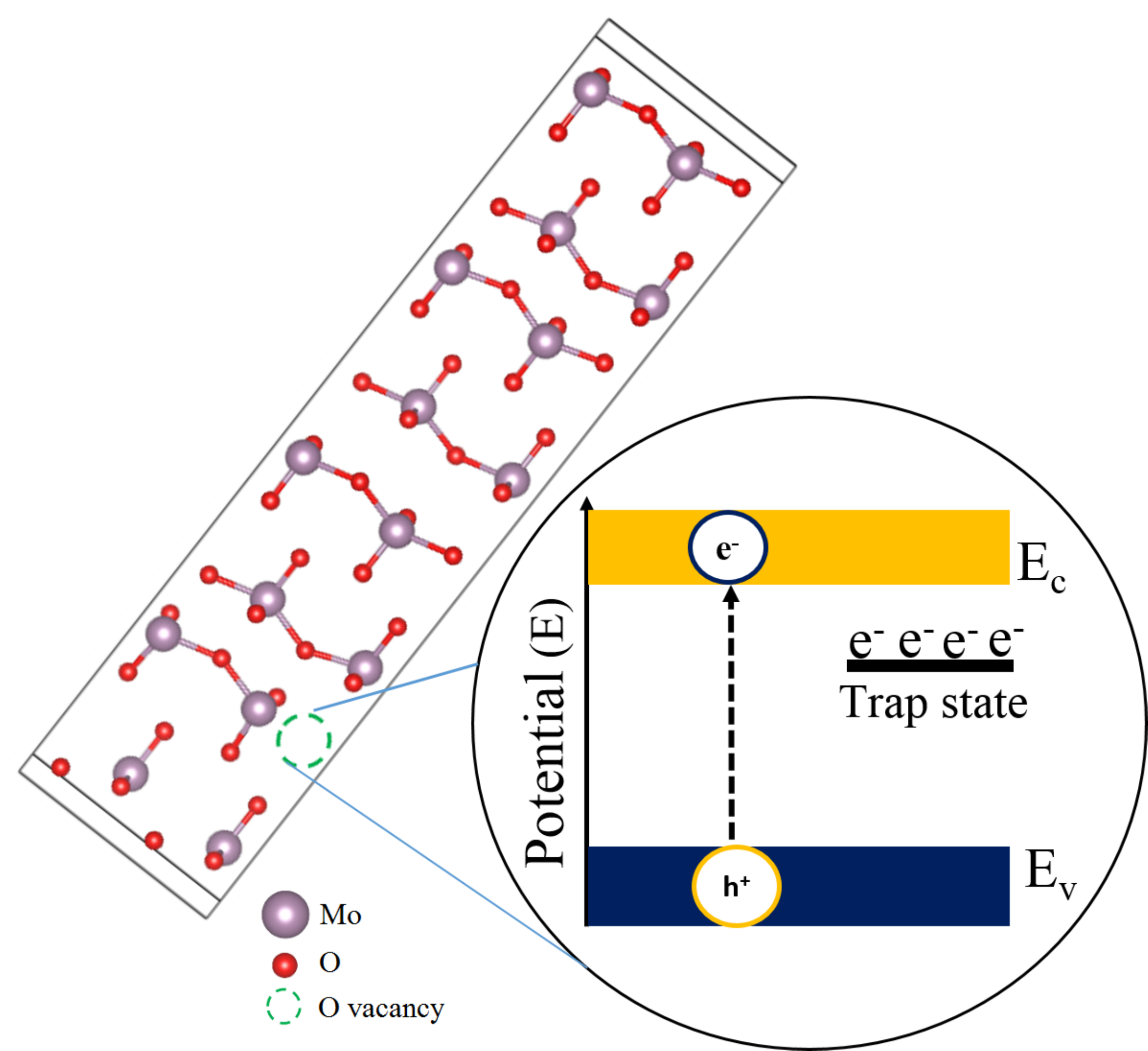}\caption{Schematic diagram showing the formation of a vacancy-induced trap
state in between the valance band maxima (VBM) and conduction band
minima (CBM).}
\label{fig:Fig6.1} 
\end{figure}

\begin{figure}[ht]
\includegraphics[width=1\linewidth]{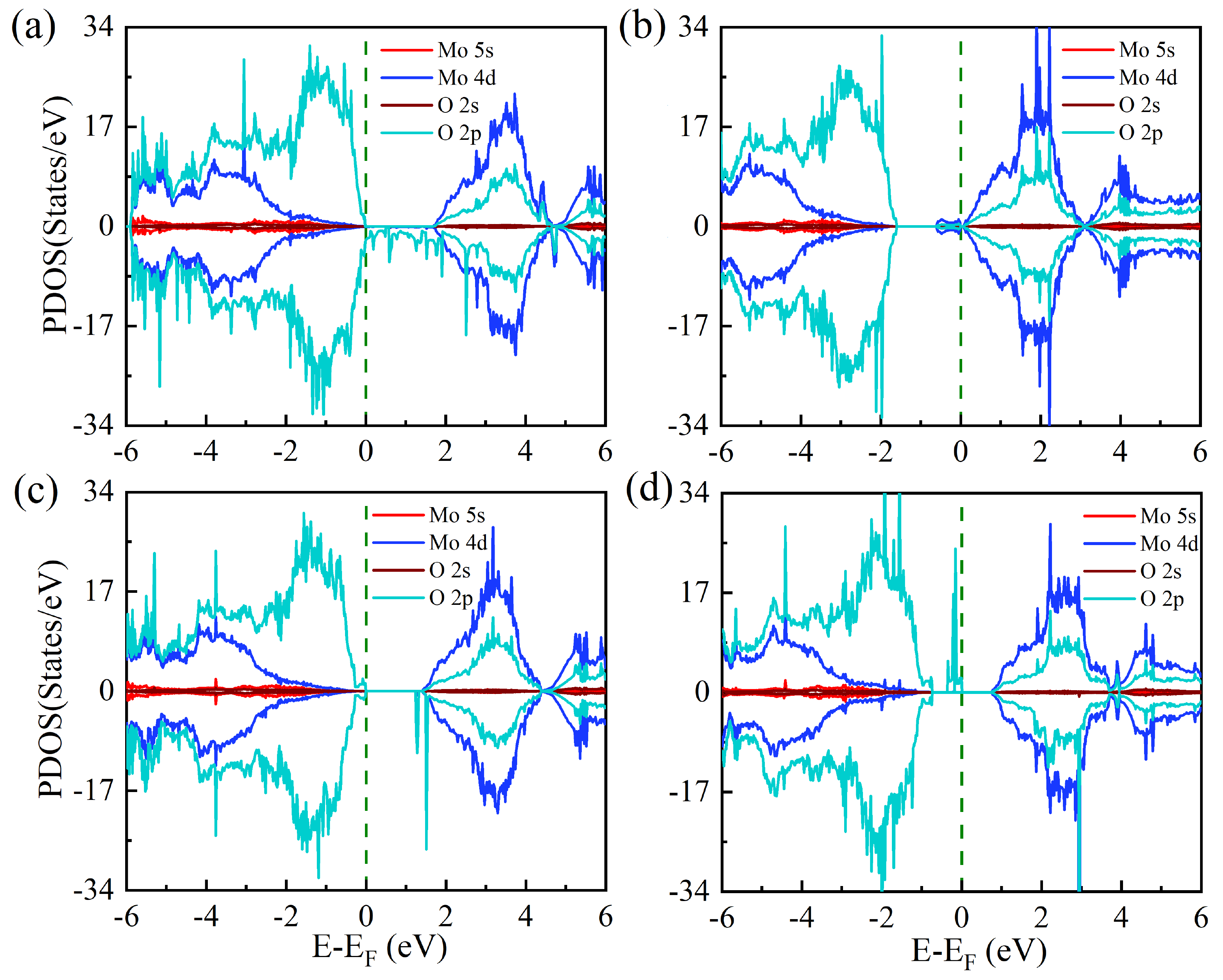}\caption{Partial density of states (PDOS) of Mo (5$s$ and 4$d$ orbitals)
and O (2$s$ and 2$p$ orbitals) in $\alpha$-MoO$_{3}$ $2\times2\times1$
supercell computed using the GGA with (a) Mo vacancy, (b) O vacancy,
and for (c,d) Mo-O co-vacancies of types P1 and P2, respectively.}
\label{fig:Fig7.1} 
\end{figure}

In Figs.~\ref{fig:Fig4.1}(a)-\ref{fig:Fig4.1}(d), we present the
TDOS corresponding to the cases with Mo, O, and Mo-O co-vacancies
of types P1 and P2, respectively, using GGA. It is obvious from the figures that
the electronic properties of the material change significantly after
the introduction of vacancies leading to the appearance of the mid-gap
states. With a Mo vacancy (see Fig.\ref{fig:Fig4.1}(a)) that corresponds to 6.25 at$\%$ of vacancy concentration, the system
becomes half-metallic with a band gap of 1.63 eV in the majority spin
states, while minority spin states show metallic behavior. For the
case of O vacancy (see Fig.\ref{fig:Fig4.1}(b)) that corresponds to 2.08 at$\%$ of vacancy concentration, the band gap disappears
completely due to the emergence of mid-gap states. Previously, Noby~\emph{et al.} experimentally showed the impact of O vacancy on $\alpha$-MoO$_{3}$ by considering different oxidizing (oxygen gain) and reducing (oxygen loss) atmospheres~\cite{noby2022oxygen}. They prepared the samples in different atmospheres (i.e., as prepared, O$_2$-treated, H$_2$-treated, N$_2$-treated and vacuum-treated) and found that the increase in the O vacancies/deficiency led to an increase in electrical conductivity, i.e., a transition from the semi-insulating to the conducting behavior. Chiam~\emph{et al.} investigated the defects-stability in $\alpha$-MoO$_{3}$ and its role in organic solar cells~\cite{chiam2012investigating}. They also investigated the presence of mixed oxidation states of Mo (i.e., 4$^{+}$, 5$^{+}$ and 6$^{+}$); as a result of this, they concluded that the defects/vacancies play an important role in the enhancement of device performance. In the case of both types of Mo–O co-vacancies, different band gaps are obtained. A reduced band gap is found in both the P1 and P2 co-vacancies.
With the P1 (P2) type of co-vacancy, a band gap of 1.39 (0.78) eV
is computed due to the majority spin-states, while minority spin-states
show a band gap of 1.23 (1.47) eV. Further, Figs.~\ref{fig:Fig5.1}(a)-~\ref{fig:Fig5.1}(h)
show the band structure for the spin-up and spin-down states corresponding
to single Mo, O, and Mo-O co-vacancies of P1 and P2 types in the $2\times2\times1$
supercell, respectively. For O vacancy (see Figs.~\ref{fig:Fig5.1}(c)
and ~\ref{fig:Fig5.1}(d)), the band structure is identical for the
spin-up and spin-down states, which signifies that the system remains
non-magnetic in the case of O vacancy. For the Mo vacancy (see Fig.~\ref{fig:Fig5.1}(a)
and ~\ref{fig:Fig5.1}(b)), a magnetic moment of 5.98~$\mu_{B}$ is
obtained, while in the case of Mo-O co-vacancies (see Fig.~\ref{fig:Fig5.1}(e)-
\ref{fig:Fig5.1}(h) irrespective of their types (i.e., P1 and P2),
a magnetic moment of 2.0~$\mu_{B}$ is computed.
Here, we also highlight the fact that our obtained results are valid for the Mo vacancy concentration of 6.25 at$\%$ and an O vacancy concentration of 2.08 at$\%$. However, in dilute defect limits, the vacancy-induced states, in all likelihood, will be much more localized, as a result of which a single vacancy may not be able to change the transport behaviour of samples from semiconducting to half-metallic/metallic.
In Fig. \ref{fig:Fig6.1}, through a schematic, we demonstrate the formation of an O-vacancy-induced mid-gap state between the VBM and CBM, which is nothing but a trap state. The trap states have additional electrons that modify the magneto-optoelectronic properties of the defective system, as compared to the pristine one. 
In order to achieve a deeper understanding of the nature of mid-gap
states, we computed the PDOS, and our results are presented in Fig.
\ref{fig:Fig7.1}. It is clear from the PDOS plot that for the case
of Mo vacancy (Fig.~\ref{fig:Fig7.1}(a) and Mo-O co-vacancies of
both types ((Figs.~\ref{fig:Fig7.1}(c) and \ref{fig:Fig7.1}(d)),
the contribution to the defect states is more from the O atoms, while
for the case of O vacancy (Fig. \ref{fig:Fig7.1}(b), Mo atoms contribute
remarkably to the defect states.

\begin{figure}[b]
\includegraphics[width=1\linewidth]{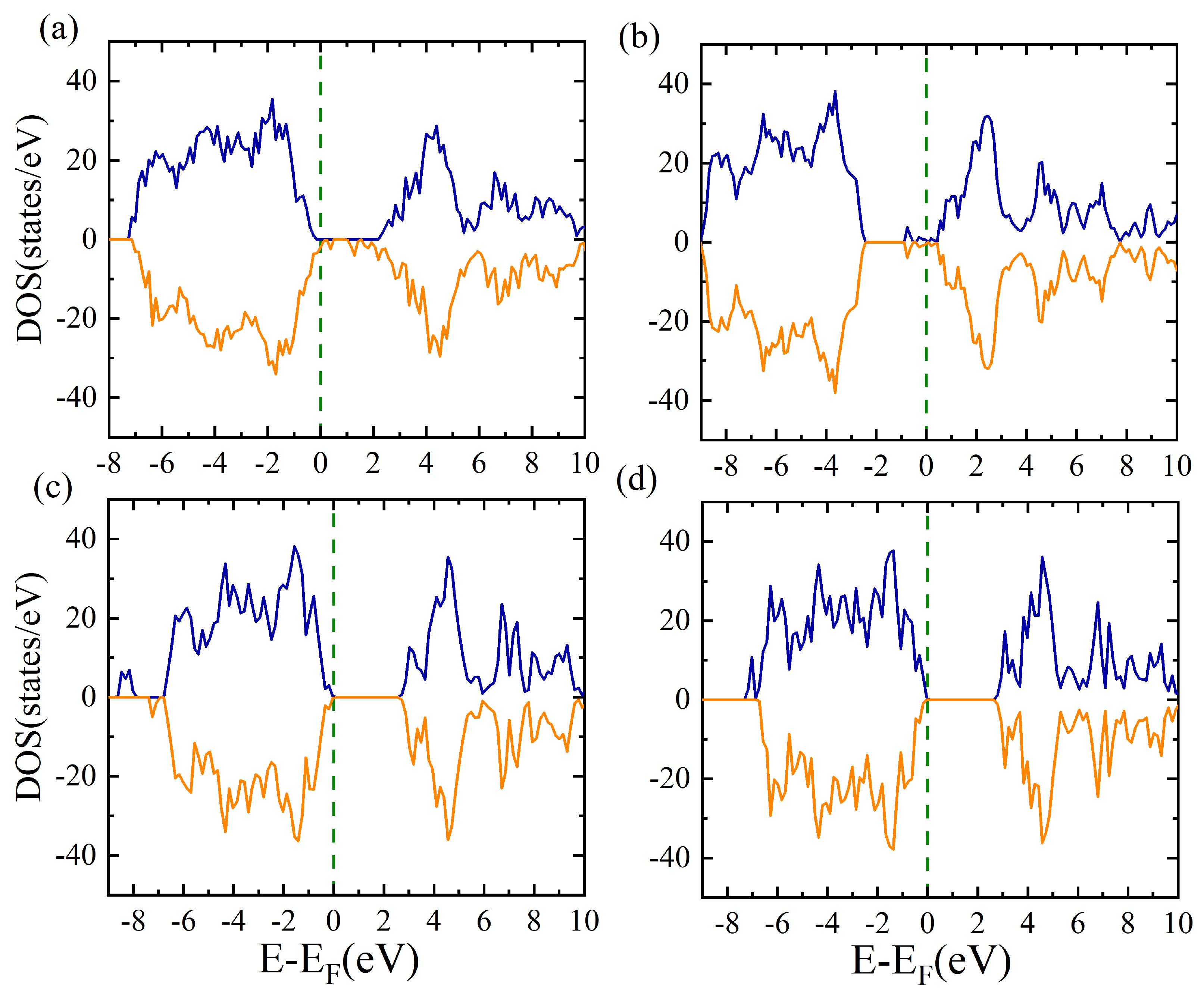}\caption{The total density states of $\alpha$-MoO$_{3}$ ($2\times2\times1$ supercell) with (a) Mo vacancy, (b) O vacancy, and Mo-O co-vacancies of types (c) P1 and (d) P2, respectively, using HSE06.}
\label{fig:hsedos} 
\end{figure}

Next, we repeated our defect calculations with the hybrid functional, i.e., HSE06, in order to verify our GGA-based results. In the MO-vacancy case (see Fig.~\ref{fig:hsedos}(a)), our HSE06 calculations reveal the half-metallic nature of the system, as obtained in our GGA results as well. While minority spin states exhibit metallic behavior, the band gap in the majority spin states is computed to be 2.33 eV, which is higher than the GGA computed band gap of 1.63 eV. Moreover, a magnetic moment of 6.0 ~$\mu_{B}$ is obtained that matches well with our GGA computed magnetic moment of 5.98~$\mu_{B}$. In the O vacancy case (see Fig.~\ref{fig:hsedos}(b)), HSE06 functional predicts the system to be metallic in agreement with our GGA results. Further, the system continues to be non-magnetic; up-spin and down-spin states show symmetrical nature, supporting our GGA findings (see Fig.~\ref{fig:Fig4.1}(b)). In P1 and P2 co-vacancies (see Fig.~\ref{fig:hsedos}(c-d)), semiconducting nature is reported for both, with band gaps of 2.45 eV and 2.66 eV, respectively, which are significantly larger than the corresponding gaps obtained in the GGA calculations. A magnetic moment of 2~$\mu_{B}$ is obtained in both types of co-vacancies in HSE06 calculations, in full agreement with the GGA results.

\subsection{Spin density analysis}

\begin{figure}[ht]
\includegraphics[width=1\linewidth]{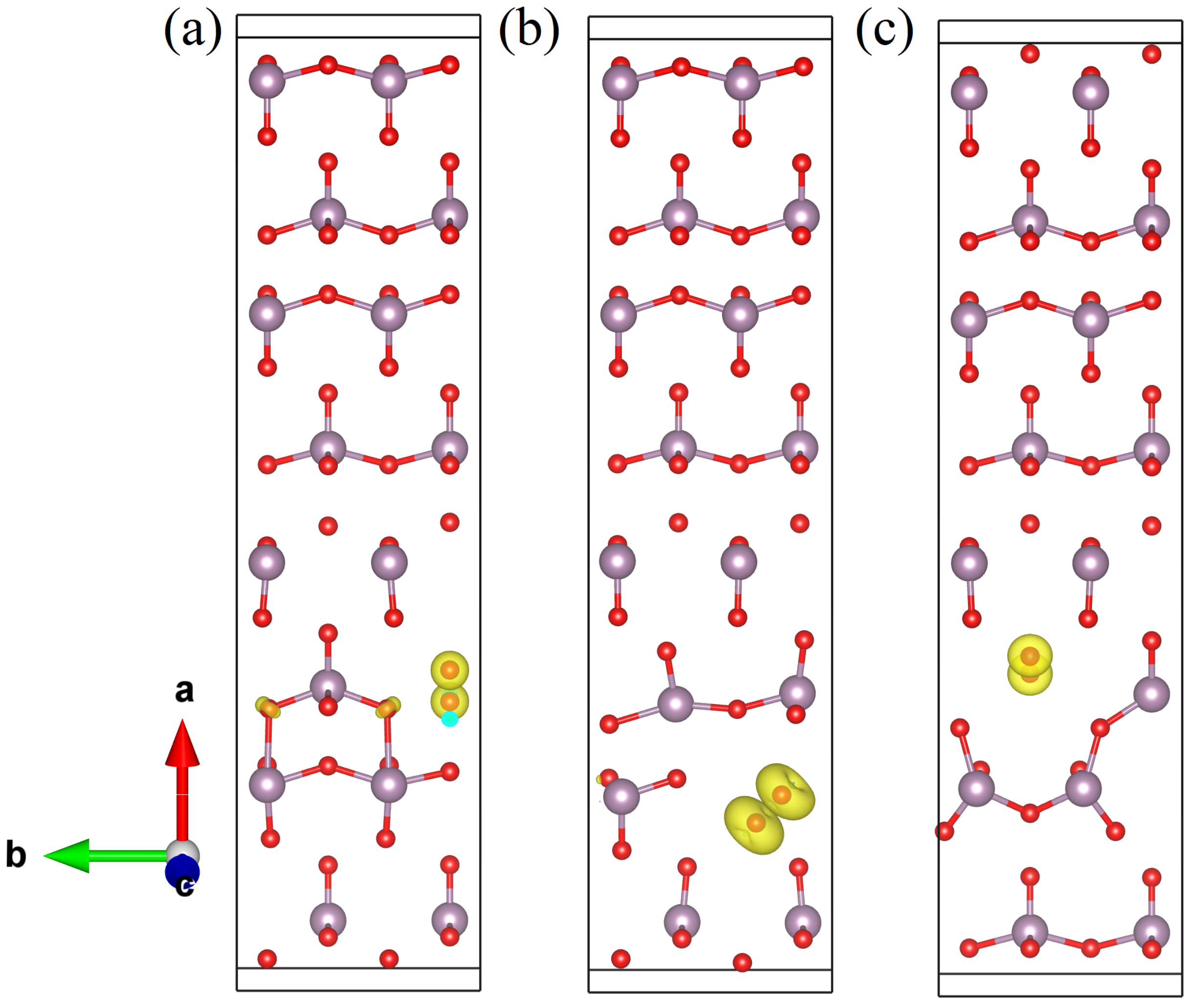}\caption{The spin density plots ($\Delta\rho$= $\rho_{\uparrow}$ - $\rho_{\downarrow}$)
of 2$\times$2$\times$1 supercell of $\alpha$-MoO$_{3}$ using GGA
with the (a) Mo vacancy for the isovalue of 0.099e and (b,c) Mo-O co-vacancies
of types P1 and P2 for the isovalue of 0.007e and 0.041e, respectively.}
\label{fig:Fig8.1} 
\end{figure}

We performed the spin density analysis to understand the contribution
of Mo and O atoms to the magnetic moment induced in different vacancy
configurations of 2$\times$2$\times$1 supercell of $\alpha$-MoO$_{3}$.
Figs.~\ref{fig:Fig8.1}(a)-~\ref{fig:Fig8.1}(c) show the spin density
plots obtained by substracting the up-spin ($\rho_{\uparrow}$) and
down-spin ($\rho_{\downarrow}$) channels of charge densities correspond
to the Mo vacancy and Mo-O co-vacancies of types P1 and P2, for the
isovalues of 0.099e, 0.007e, and 0.041e, respectively. As shown in the figures, the spin density isosurface ( shown in yellow color) is mostly
concentrated around the O atoms (nearest neighbor atoms of vacancies)
in all three cases, which reflects the fact that in all three cases,
the contribution from O atoms is more to the magnetic moment. Our
spin density plots are qualitatively compatible with the PDOS analysis
(see Fig.~\ref{fig:Fig7.1}). In addition, the spin density isosurface
is dumbbell-shaped, indicating a higher contribution of the $p$ orbitals of oxygen to the induced magnetic moments.

\subsection{Practical feasibility }

We computed the formation energy for (i) Mo vacancy, (ii) O vacancy,
and (iii) Mo-O co-vacancies of both types, i.e., P1 and P2, respectively,
for the practical realization of the systems with respect to these defect configurations.
Before discussing the formation energy of different defect configurations,
first, we calculated the formation energy for pristine $\alpha$-MoO$_{3}$
by considering metallic Mo and O$_{2}$ gas using the formula 
\begin{equation}
\begin{split} & E_{form}^{MoO_{3}}=\mu_{Mo}^{MoO_{3}}+3\mu_{O}^{MoO_{3}}-(\mu_{Mo}^{0}+3\mu_{O}^{0})\\
 & \quad\qquad=E_{MoO_{3}}-(\mu_{Mo}^{0}+3\mu_{O}^{0}).
\end{split}
\label{equ:equ6}
\end{equation}
Where $E_{MoO_{3}}$ is the total energy of the pristine $\alpha$-MoO$_{3}$
energy per formula unit, and $\mu_{Mo}^{MoO_{3}}$ and $\mu_{O}^{MoO_{3}}$
are the chemical potentials of Mo and O in $\alpha$-MoO$_{3}$. Moreover,
$\mu_{Mo}^{0}$ and $\mu_{O}^{0}$ are the chemical potential of Mo
and O in pure molybdenum metal and O$_{2}$ gas, respectively. 

Using Eq.~\ref{equ:equ6}, and the obtained values in our DFT simulations, i.e., $E_{MoO_{3}}=-32.88$ eV, $\mu_{Mo}^{0}=-10.94$ eV, and $\mu_{O}^{0}=0.5\times(-9.77)=-4.885$ eV, we calculated $E_{form}^{MoO_{3}}=-7.27$ eV, which is in good agreement with the theoretically reported value of -6.77 eV~\citep{das2019structural} and the experimental value of -7.72 eV~\citep{speight2005inorganic}.
A good matching of formation energy with the previously reported theoretical and experimental results gives us confidence and shows the correctness of our simulation methods.

Next, we calculated the formation energy for the defective systems
using Eq.~\ref{eq:eq5}, as discussed previously. Using the stability
conditions, the following range of chemical potentials for Mo and
O have been derived and further used in the formation energy calculation
of the defective system 
\begin{equation}
\mu_{Mo}^{min}=\mu_{Mo}^{0}+E_{form}^{MoO_{3}}<\mu_{Mo}^{MoO_{3}}<\mu_{Mo}^{0}.\label{equ:equ8}
\end{equation}
\begin{equation}
\mu_{O}^{min}=\mu_{O}^{0}+\frac{1}{3}E_{form}^{MoO_{3}}<\mu_{O}^{MoO_{3}}<\mu_{O}^{0}.\label{equ:equ9}
\end{equation}

\begin{figure}[t]
\includegraphics[width=1\linewidth]{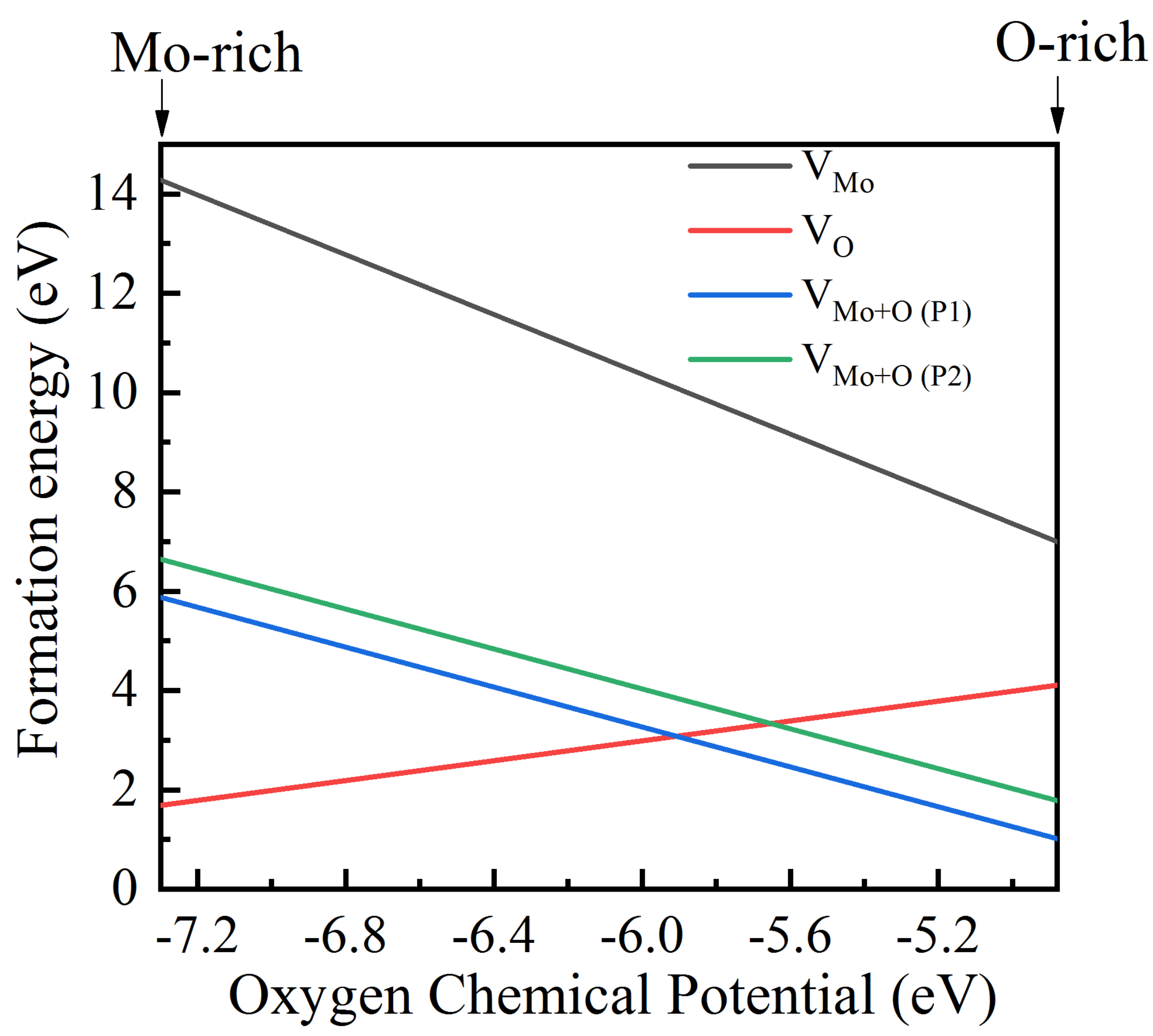}\caption{Formation energy of $\alpha$-MoO$_{3}$ corresponding to different
vacancy configurations (as shown in Fig.~\ref{fig:Fig1.1}) with
respect to O chemical potential. The left side shows the Mo-rich,
and the right side shows the O-rich limits.}
\label{fig:formation_plot}
\end{figure}

In the above equations, $\mu_{Mo}^{0}$ and $\mu_{O}^{min}$ denote
the chemical potentials of Mo and O, in Mo-rich conditions, while
$\mu_{Mo}^{min}$ and $\mu_{O}^{0}$ represent those in O-rich conditions.
The calculated value of chemical potential in Mo-rich condition for
Mo (O) is -10.94 (-7.30) eV, whereas in the O-rich condition it is
-18.22 (-4.88) eV. Our computed results of formation energy corresponding
to different vacancy configurations are presented in Table~\ref{Table:formation_table}
(Fig.~\ref{fig:formation_plot}). 

\begin{table}[ht]
\centering 
\caption{\label{tab:table1} Formation energies (eV) corresponding to different
vacancy configurations in $\alpha$-MoO$_{3}$, in both Mo-rich and
O-rich environments .}

\begin{ruledtabular}
\begin{tabular}{ccccc}
Vacancy types$\rightarrow$ & V$_{Mo}$ & V$_{O}$ & V$_{Mo+O(P1)}$ & V$_{Mo+O(P2)}$\tabularnewline
\hline 
Mo-rich & 14.28 & 1.69 & 5.58 & 6.65\tabularnewline
O-rich & 7.0 & 4.11 & 1.02 & 1.79\tabularnewline
\end{tabular}
\end{ruledtabular}

\label{Table:formation_table}
\end{table}

It is clear from Fig.~\ref{fig:formation_plot} and Table \ref{tab:table1}
that the formation energy for Mo vacancy is very high compared to
other defect configurations for the entire range of O chemical potential
indicating that Mo vacancy is difficult to form at ordinary conditions
in $\alpha$-MoO$_{3}$~\cite{peelaers2017controlling}. Further, O vacancy in the proximity of V$_{Mo}$
(i.e., Mo-O co-vacancies) reduces the formation energy of Mo vacancy
considerably. Therefore, our first-principles results indicate that
Mo--O co-vacancies support induced magnetic moment in $\alpha$-MoO$_{3}$,
along with reducing the formation energy, and thus stabilizing the
Mo vacancy that is responsible for inducing the magnetic moment in
the system, as discussed in the previous section. Further, in the
Mo-rich condition (i.e., O-poor), O vacancy has the lowest formation
energy of 1.69 eV. Further, for the sake of comparison, we note that Cao~\emph{et
al.} reported $\sim$ 1.4 eV of formation energy for S vacancy in
MoS$_{2}$ under Mo-rich condition~\citep{cao2016first}, while Kuklin~\emph{et
al.} reported 1.64 eV for Se vacancy in PdSe$_{2}$ under Pd-rich
conditions~\citep{kuklin2021point}. Our computed value of formation
energy for O vacancy is within the usual range of formation energies
reported for transition-metal dichalcogenide systems in the literature
With the increase in O chemical potential, i.e., towards the O-rich
limit, Mo-O co-vacancies of both types start dominating and are more
stable compared to the O vacancy. Moreover, the P2 type of Mo-O co-vacancy
shows 0.77 eV higher formation energy than P1, implying that the P1
type of co-vacancy is more stable, but on the other hand, both types
of co-vacancies exhibit lower formation energy than O vacancy. Here
we would like to emphasize the fact that experimentally the stoichiometry
of the growth circumstances plays a very crucial role and may affect
formation energy values significantly~\citep{koos2019influence}.

\begin{figure}[ht]
\includegraphics[width=1\linewidth]{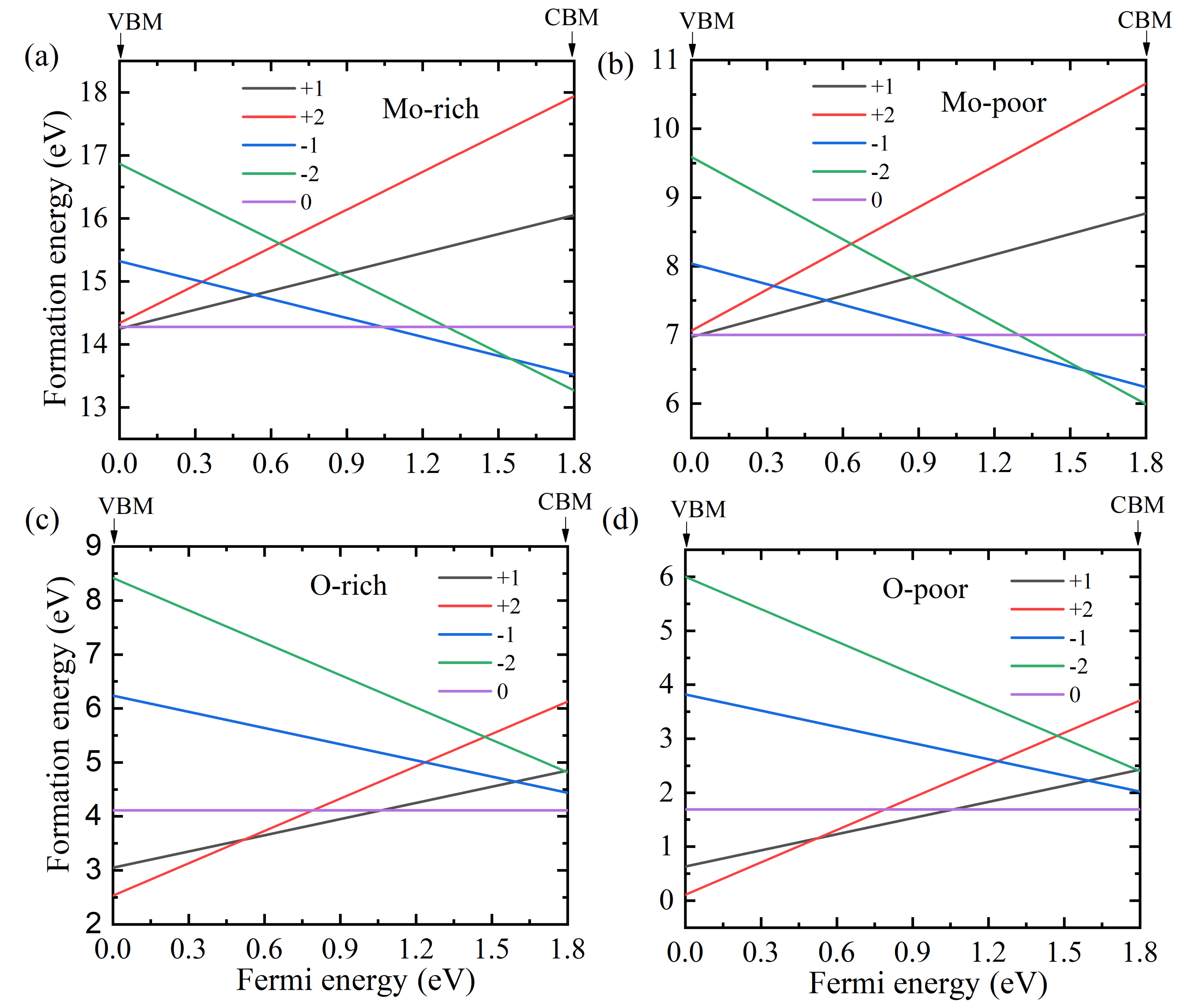}\caption{Charged defects formation energies for the (a,b) Mo vacancy and (c,d) O vacancy in Mo-rich/poor and O-rich/poor conditions as a function of the Fermi energy.}
\label{fig:chagre_formation} 
\end{figure}

Till now, we discussed the formation energy of neutral vacancies. Next, we considered different charged defect states. We considered +1, +2, -1 and -2 charge states for both Mo and O vacancies and corresponding formation energies are shown in Fig.~\ref{fig:chagre_formation}. From the formation energy graphs, it is clear that, as in neutral states, the formation energy of Mo vacancy in different charged states is still high compared with O vacancy. Further, in the O-vacancy case, in both rich and poor conditions, the +1 and +2 states are more stable compared to the negatively charged states, as reported previously~\cite{peelaers2017controlling}. This clearly indicates that the O-vacancy donates electrons to the conduction band to form a stable state. Therefore, O vacancy behaves like a shallow donor, as shown by Noby~\emph{et al.}~\cite{noby2022oxygen} and is responsible for the unintentional n-type semiconducting behavior of $\alpha$-MoO$_{3}$.

\subsection{Ab-initio MD simulations }

In this section, we discuss the results obtained from Ab-initio molecular
dynamics (AIMD) simulations, which are performed to investigate the
stability and sustainability of the induced magnetic moment, and also
to verify the stability of the structure, at room temperature. For
technological applications, such as for implementation in spintronic
devices, it is important to check not only the structural stability,
but also the survival of induced magnetic moments at room temperature.
In our study, O vacancy does not induce any kind of magnetic moment
in the system, therefore, we performed the AIMD simulations for Mo-O
co-vacancy of type P1 owing to the fact that although both types of
co-vacancies induce the same magnetic moment (i.e., 2~$\mu_{B}$),
the P1 type of co-vacancy is found to be more stable, as discussed
earlier. 
\begin{figure*}
\includegraphics[width=1\linewidth]{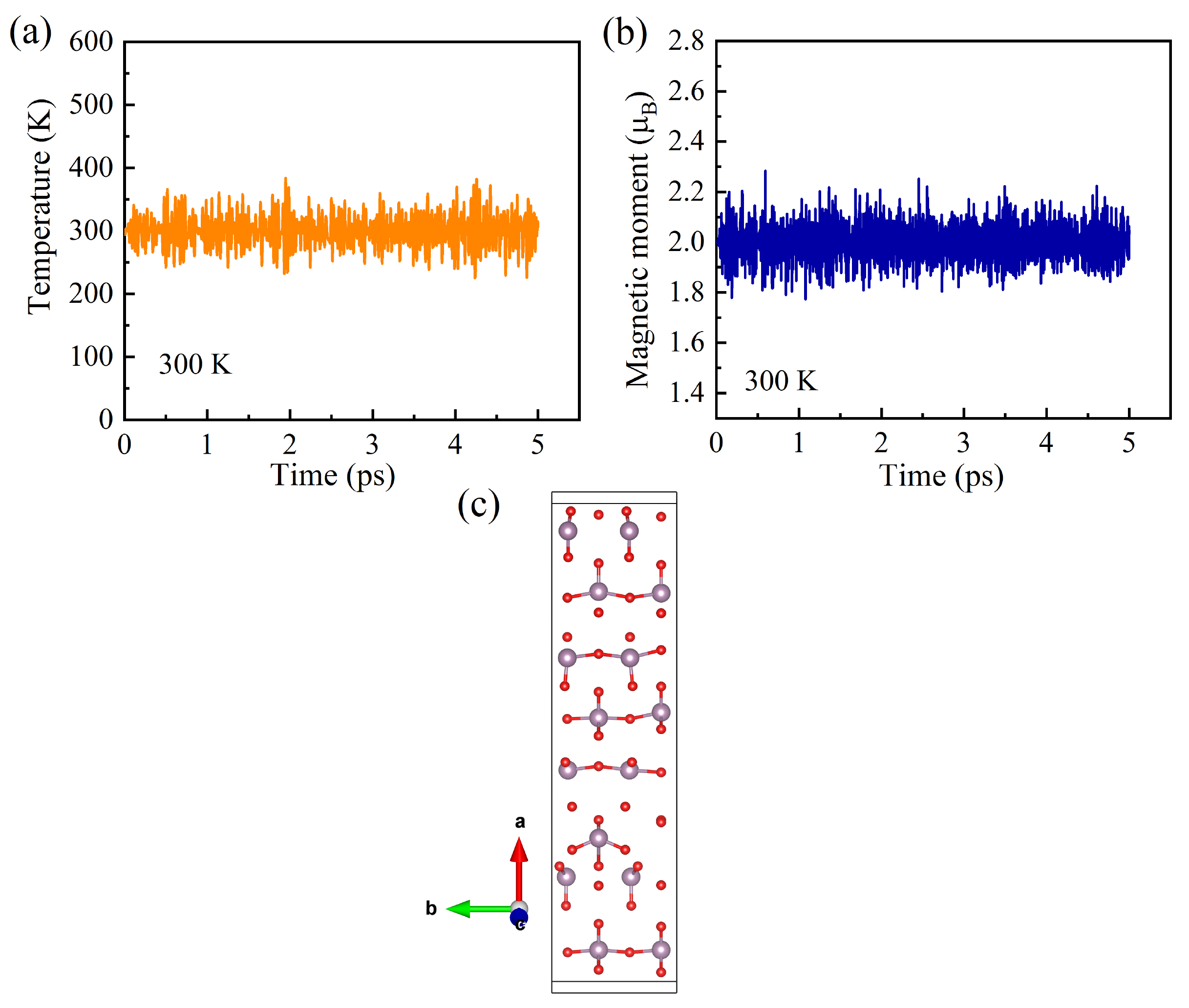}\caption{Variation of (a) temperature and (b) magnetic moment as a function
of AIMD simulation steps at the initial temperature 300 K. (c) Mo-O
co-vacancy structure of type P1 obtained at 300 K.}
\label{fig:md}
\end{figure*}

Experimentally, to favor the P1 type of Mo-O co-vacancy, one can tune
the O chemical potential ($\mu_{O}$) within the stability limits
as described in Eq.~\ref{equ:equ9} by using the equation

\begin{equation}
\mu_{O}(T,P)=\mu_{O}(T,P^{0})+\frac{1}{2}KT\ln(\frac{P}{P^{0}}).
\end{equation}
Here, $T$ and $P$ denote the temperature and pressure of O$_{2}$,
$K$ denotes Boltzmann's constant, and $P^{0}$ = 1 atm.

We performed the AIMD simulations for the time period of 5 ps, with the
initial temperature fixed at the room temperature taken to be 300
K. Figs.~\ref{fig:md}(a) and \ref{fig:md}(b) show the variation
of temperature and magnetic moment as a function of time. In Fig.~\ref{fig:md}(c),
we show the Mo-O co-vacancy structure of P1 type obtained at 300 K.
After a time period of 5 ps, we found that the structure remains stable
(Fig.~\ref{fig:md}(c)) with an average magnetic moment of 2~$\mu_{B}$,
which exactly matches the obtained value of magnetic moment in our
DFT results. The induced magnetic moment is found to be very stable,
with minimal variations around 2~$\mu_{B}$, and sustained at the room
temperature ( Fig.~\ref{fig:md}(a)-~\ref{fig:md}(b)).

\subsection{Optical Absorption Spectra }
\begin{figure}[ht]
\includegraphics[width=1\linewidth]{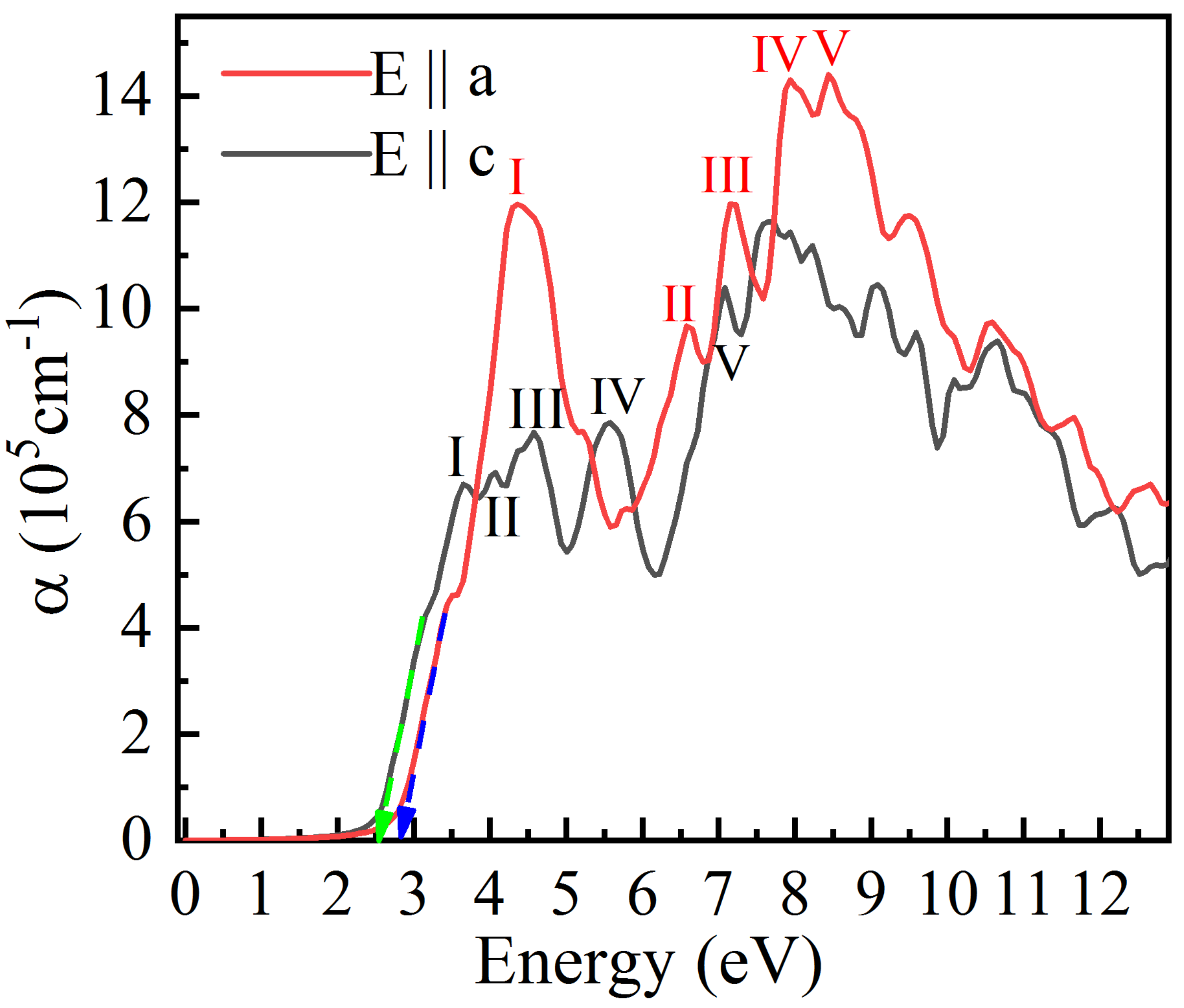}\caption{Our calculated absorption spectra of 2$\times$2$\times$1 supercell
of pristine $\alpha$-MoO$_{3}$ for E$\parallel$a (red color) and
E$\parallel$c (black color) using the first-principles DFT, employing
PBE functional.}
\label{fig:Fig9.1} 
\end{figure}

Now, we analyze the optical properties of the system because the optical
characteristics of semiconductor oxide materials have generated tremendous
research interest due to their potential use in solar cells, photovoltaic
materials,~\emph{etc.} In Fig.~\ref{fig:Fig9.1}, we present the
optical absorption spectra of pristine $\alpha$-MoO$_{3}$ using
the GGA approach. Huang~\emph{et al.} performed a comprehensive first-principles
DFT study of the electronic structure and optical properties of pristine
and H-doped $\alpha$-MoO$_{3}$ using the optB88 functional, and
reported an optical gap $\approx$ 2.0 eV~\citep{huang2014impact}.
Experimentally, Itoh~\emph{et al.} measured the optical gap of the
single crystal $\alpha$-MoO$_{3}$ to be 3.5 eV~\citep{Itoh_2001},
while in the thin film samples, the corresponding value was found
to be 3.1 eV~\citep{carcia1987synthesis,sabhapathi1994structural,sabhapathi1995growth}. In our study, we used the convention of Itoh~\emph{et al.} for the a and c axes
while plotting the optical absorption spectra~\citep{Itoh_2001}.

\begin{figure}[t]
\includegraphics[width=1\linewidth]{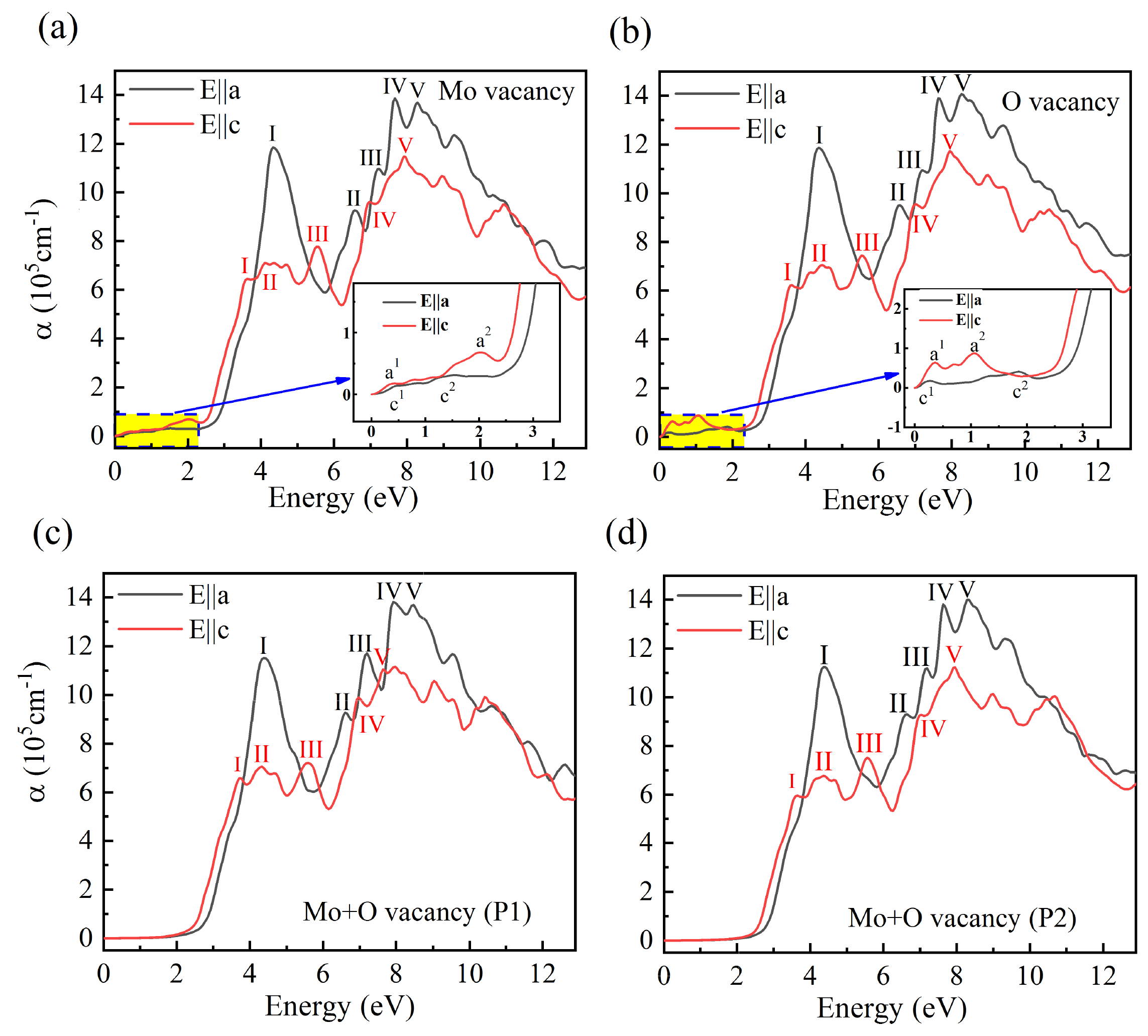}\caption{Absorption spectra of 2$\times$2$\times$1 supercell of $\alpha$-MoO$_{3}$
for {E$\parallel$ a} (red color) and {E$\parallel$ c} (black
color) with (a) Mo-vacancy, (b) O-vacancy, and (c,d) Mo-O co-vacancies
of types P1 and P2, respectively. (black color).}
\label{fig:Fig10.1} 
\end{figure}

From Fig.~\ref{fig:Fig9.1} it is obvious that the optical gaps (the
energy marking the onset of optical absorption) obtained in our PBE
functional-based calculations for E$\parallel$a (E$\parallel$c)
are 2.93 (2.60) eV, which are 0.17 (0.5) eV lower than the thin-film
values, and 0.57 (0.9) eV below the single-crystal values. Our DFT computed absorption spectra show considerable anisotropic
behavior with respect to the polarization (i.e., E$\parallel$a vs.
E$\parallel$c), as reported previously in single crystal and thin
film-based experiments\citep{Itoh_2001,carcia1987synthesis,sabhapathi1994structural,sabhapathi1995growth}.
\begin{table*}[t]
\caption{Calculated peak energy values (eV) from absorption spectra of 2$\times$2$\times$1
supercell of pristine $\alpha$-MoO$_{3}$. Inside the parentheses,
orbitals that contribute to the single electron transition to that
obtained peak value are mentioned. $v$ and $c$ indicate the VBM
and CBM, and c+n indicates the nth conduction band.}

\begin{ruledtabular}
\begin{tabular}{c|c|cc}
Peak No.  & \multicolumn{2}{c}{ Peak energy values (eV) } & \tabularnewline
\hline 
 & E$\parallel$a  & E$\parallel$c  & \tabularnewline
\hline 
I  & 4.36 ($|v\rightarrow c+6\rangle$)  & 3.64 ($|v\rightarrow c+4\rangle$)  & \tabularnewline
II  & 6.58 ($|v\rightarrow c+16\rangle$)  & 4.07 ($|v\rightarrow c+4\rangle$)  & \tabularnewline
III  & 7.15 ($|v\rightarrow c+20\rangle$)  & 4.57 ($|v\rightarrow c+7\rangle$)  & \tabularnewline
IV  & 7.94 ($|v\rightarrow c+18\rangle$)  & 5.57 ($|v\rightarrow c+12\rangle$)  & \tabularnewline
V  & 8.44 ($|v\rightarrow c+22\rangle$)  & 7.08 ($|v\rightarrow c+17\rangle$)  & \tabularnewline
\end{tabular}\label{tab:tablei} 
\end{ruledtabular}

\end{table*}
Moreover, in E$\parallel$a component of the spectrum, we obtained
five different peaks at 4.36 eV, 6.58 eV, 7.22 eV, 7.94 eV, and 8.44
eV. On the other hand, in E$\parallel$c component, the peaks are
at 3.64 eV, 4.07 eV, 4.57 eV, 5.57 eV, and 7.08 eV, respectively.
The obtained peak energy values in the absorption spectra for E$\parallel$a
and E$\parallel$c are shown in Table~\ref{tab:tablei}; inside the
parentheses, we show the involved Bloch orbitals that contribute to
the single-electron transition to that peak value. We found that for
the E$\parallel$a component, all the transitions for five different
peak values occur at the $\Gamma$ point, whereas in E$\parallel$c, corresponding to the first and the third peak, the transitions
are at the T point, while for the rest of the peaks, i.e., the second,
fourth, and fifth, the occur at the $\Gamma$ point. Although, we have not computed the excitonic effects in the absorption spectrum, however, we expect them to be significant, because the exciton binding energy in MoO$_3$ has been estimated to be in the range of 21-28 meV~\cite{shahrokhi2020understanding,fogle1972semiconductor}
\begin{table*}[t]
\caption{Calculated peak energy values (eV) from absorption spectra corresponding
to different vacancy configurations of 2$\times$2$\times$1 supercell
of $\alpha$-MoO$_{3}$.}

\begin{ruledtabular}
\begin{tabular}{c|c|c|c|c|c|c|c|ccc}
 & \multicolumn{4}{c|}{ {E$\parallel$a} (eV) } & \multicolumn{4}{c}{ {E$\parallel$c} (eV) } &  & \tabularnewline
\hline 
Peak No.  & Mo  & O  & Mo-O co-  & Mo-O co-  & Mo  & O  & Mo-O co-  & Mo-O co-  &  & \tabularnewline
 & vacancy  & vacancy & vacancy (P1) & vacancy (P2)  & vacancy  & vacancy & vacancy (P1) & vacancy (P2)  &  & \tabularnewline
\hline 
I  & 4.36  & 4.37  & 4.35  & 4.33  & 3.63  & 3.71  & 3.59  & 3.62  &  & \tabularnewline
II  & 6.58  & 6.59  & 6.64  & 6.57  & 4.45  & 4.32  & 4.35  & 4.09  &  & \tabularnewline
III  & 7.17  & 7.20  & 7.19  & 7.24  & 5.54  & 5.59  & 5.55  & 5.52  &  & \tabularnewline
IV  & 7.63  & 7.92  & 7.62  & 7.67  & 6.69  & 6.98  & 6.97  & 6.95  &  & \tabularnewline
V  & 8.26  & 8.47  & 8.28  & 8.29  & 7.94  & 7.64  & 7.95  & 7.95  &  & \tabularnewline
\end{tabular}\label{tab:table2} 
\end{ruledtabular}

\end{table*}
Next, we analyze the optical properties of defective systems of 2$\times$2$\times$1
supercell of $\alpha$-MoO$_{3}$. Figs.~\ref{fig:Fig10.1}(a)-~\ref{fig:Fig10.1}(d)
show the absorption spectra of the supercell with Mo-vacancy, O-vacancy,
and Mo-O co-vacancies of types P1 ad P2, respectively. It is obvious
from the plots that due to the introduction of vacancies into the
system, some new peaks have emerged in the absorption spectra below
the main band edge. In the Mo vacancy case (Fig.~\ref{fig:Fig10.1}(a)),
below the band edge, two peaks are obtained at 0.47 eV (c$^{1}$)
and 1.23 eV (c$^{2}$) for the E$\parallel$a component, and 0.33
eV (a$^{1}$) and 2.0 eV (a$^{2}$) in the E$\parallel$c component
(see inset of Fig.~\ref{fig:Fig10.1}(a)). In case of the O vacancy
(Fig.~\ref{fig:Fig10.1}(b)), the peak positions below the band edge
are obtained at 0.27 eV (c$^{1}$) and 1.86 eV (c$^{2}$) for E$\parallel$a,
and 0.36 eV (a$^{1}$) and 1.04 eV (a$^{2}$) for E$\parallel$c (see
inset of Fig.~\ref{fig:Fig10.1}(b)), respectively. Interestingly,
there are no peaks obtained below the band edge in the Mo-O co-vacancies
of both types. As in the case of the pristine system, for the defective
systems also we obtain five different absorption peaks in the absorption
spectra corresponding to the Mo, O, and Mo-O co-vacancies ( Fig. \ref{fig:Fig10.1}(a)-\ref{fig:Fig10.1}(d)).
The calculated peak energy values for different defect configurations
are presented in Table \ref{tab:table2}.


\section{Conclusion}

In this work we presented a first-principles DFT-based systematic
study of electronic, optical, and magnetic properties of crystalline
$\alpha$-MoO$_{3}$ both in its pristine form and with various vacancy
configurations. Due to the vacancies, mid-gap states appear, modifying
the properties of the system significantly. From the formation energy
calculations, we conclude that O vacancies (in the Mo-rich limit)
and Mo-O co-vacancies (in the O-rich limit) are more favorable defects
in the $\alpha$-MoO$_{3}$, compared with the Mo vacancies in the
system. Although, the pristine $\alpha$-MoO$_{3}$ is a non-magnetic
semiconductor, however, our calculations predict that configurations
with Mo vacancy and Mo-O co-vacancies give rise to a finite magnetic
moment in the system. Mo vacancy leads not only to a very large magnetic
moment of 5.98~$\mu_{B}$, but also gives rise to half-metallicity
in the system, which can be useful in spintronic applications. In the
case of Mo-O co-vacancies, a comparatively lower but still significant
magnetic moment of 2.0~$\mu_{B}$ is induced in the system, indicating the possibility of the realization of magnetic behavior in the material by defect engineering. In the optical absorption spectra
of configurations Mo/O vacancies, we observe peaks in the gap region,
however, these peaks vanish for configurations with Mo-O co-vacancies
just like in the pristine state of $\alpha$-MoO$_{3}$. This suggests
that optical absorption spectroscopy can be used for defect identification
in the system.


\section{Acknowledgement}
P S acknowledges UGC, India, for the senior research fellowship [Grant No. 1330/(CSIR-UGC NET JUNE 2018)]. All the calculation results were obtained using the computational facilities (Spacetime cluster) of the Department of Physics, IIT Bombay.

\bibliographystyle{apsrev4-2}
\bibliography{MoO3}

\end{document}